\documentclass[12pt]{article}
\usepackage{amsfonts}
\usepackage[utf8]{inputenc}
\usepackage[english]{babel}
\usepackage{bm}
\usepackage{amsmath}
\usepackage{amssymb}
\usepackage{epsfig}
\usepackage{hyperref}
\usepackage{xcolor}
\usepackage{cite}

\newcommand{\bmat}{\left(\begin{array}}
	\newcommand{\emat}{\end{array}\right)}

\def\gtrsim{\mathrel{\raise.3ex\hbox{$>$\kern-.75em\lower1ex\hbox{$\sim$}}}}

\def\ap{\alpha^{\prime}}

\def\ov{\overline}
\def\un{\underline}

\def\-{\hphantom{-}}
\def\ov{\overline}
\def\s2{\frac{1}{\sqrt2}}

\def\mg{m_{3/2}}
\def\mg2{m^2_{3/2}}

\def\Dsl{\,\raise.15ex\hbox{/}\mkern-13.5mu D} 

\def\be{\begin{equation}}
	\def\ee{\end{equation}}
\def\bea{\begin{eqnarray}}
	\def\eea{\end{eqnarray}}

\newcommand{\nn}{\nonumber}

\begin{document}
	
	
	\pagestyle{plain}

	\makeatletter
	\@addtoreset{equation}{section}
	\makeatother
	\renewcommand{\theequation}{\thesection.\arabic{equation}}
	\pagestyle{empty}
	\begin{center}
		\ \
		
		\vskip .5cm
		
		\LARGE{\LARGE\bf Higher-Derivative Corrections via the Double Copy Procedure}
\vskip 0.3cm

\large{Eric Lescano$^{*}$ and Jesus A. Rodriguez$^\dag$$^\#$ 
	\\[3mm]}

{$^*$\small University of Wroclaw, Faculty of Physics and Astronomy,  \\ [-10pt]} 
{\small\it Maksa Borna 9, 50-204 Wroclaw,
Poland}

\vspace{0.3cm}

{$^\dag$\small Universidad de Buenos Aires, FCEyN, Departamento de Física, \\ [-10pt]}
{\small\it Ciudad Universitaria, 1428 Buenos Aires, Argentina}

\vspace{0.3cm}

{$^\#$\small Universidad Argentina de la Empresa (UADE), Facultad de Ingeniería y Ciencias Exactas,\\ [-10pt]}
{\small\it Lima 717, Buenos Aires, Argentina}

{\small \verb"eric.lescano@uwr.edu.pl, jarodriguez@df.uba.ar"}\\[0.5cm]
\small{\bf Abstract} \\[0.5cm]\end{center}

Recent advances in the off-shell formulation of the Double Copy (DC) procedure have revealed a profound connection between gauge theories and T-duality invariant frameworks. The main example is Double Field Theory (DFT), emerging as the the off-shell DC of Yang-Mills theory up to cubic order in perturbations. Extending this procedure to a higher-derivative gauge theory gives rise to a Higher-Derivative Double Theory (HDDT), which incorporates Weyl gravity along with $b$-field and dilaton contributions, all in a T-duality invariant manner. In this work, we show that the quadratic contributions of HDDT are directly related (up to field redefinitions) to DFT+, a T-duality invariant model associated with the bosonic string that incorporates first-order $\alpha'$ corrections upon parameterization. Our results expand the potential applications of the off-shell DC program towards constructing perturbative $\alpha'$-corrected Lagrangians, while also opening up possibilities for reversing the map by considering the single and zeroth copies.

\newpage

\setcounter{page}{1}
\pagestyle{plain}
\renewcommand{\thefootnote}{\arabic{footnote}}
\setcounter{footnote}{0}

\tableofcontents
\newpage

\section{Introduction}
The NS-NS sector of the low energy limit of string theory exhibits a global ${\rm{O}}(n,n)$ symmetry when the fields are independent of $n$ spatial coordinates \cite{ms}. This continuous T-duality symmetry is exact to all orders in $\ap$ \cite{sen}, which motivates the study of higher-order corrections in theories with manifest duality invariance, such as Double Field Theory (DFT) \cite{DFT1,DFT2}, a proposal to include T-duality as a fundamental symmetry of a field theory on a doubled space\footnote{For reviews, see \cite{DFTreview1,DFTreview2} and the second lecture of \cite{Electure}.}. DFT reformulates supergravity in terms of ${\rm{O}}(D,D)$ multiplets within a doubled geometrical framework, where $D$ is the space-time dimension, and $D\geq n$. Although the geometric structure of DFT makes it difficult to directly construct invariant objects that are quadratic in curvatures, substantial progress has been made in obtaining higher-derivative corrections \cite{HSZ, HSZ2, Hohm:2014xsa, MN, odd, gbdr1, ehs, chm, gbdr2, gbdr3, gbdr4}. In \cite{gbdr1}, an exact mechanism was introduced through a generalization of the Green-Schwarz transformation, which requires an infinite tower of ${\rm{O}}(D,D)$-covariant higher-derivative terms in the gauge-invariant action. Specifically, the first-order corrections reduce to the bi-parametric theory presented in \cite{MN}, which interpolates between different low-energy limits of String Theory. 

However, as stated in \cite{wulff}, the current form of introducing higher-derivative terms in DFT  cannot reproduce all the higher-order terms in the $\ap$ expansion of string theory and hence, developing new techniques to obtain higher-derivative terms remains a major challenge. It is within this context that the Double Copy (DC) construction \cite{Dcopy1, Dcopy2, Dcopy3, Dcopy4, Dcopy5, class1, class2, class3, class4} emerges as a promising approach to potentially provide new insights into systematically generating higher-derivative corrections. The quadratic and cubic contributions to perturbative DFT can be derived by applying an off-shell DC map to the quadratic and cubic Yang-Mills (YM) Lagrangian \cite{HohmDC}. This connection suggests that DFT is part of a broader class of T-duality invariant theories, all of which can be derived through the off-shell DC procedure applied to different gauge theories. More recent research \cite{LMR, EA} delved into a new DC prescription applied to the higher-derivative gauge theory \cite{DFJ1, DFJ2}. Particularly, in \cite{EA}, the authors studied the off-shell double copy realization of the following Lagrangian,
\bea
{\cal L} & = & a_{1}\kappa_{ab}D_{\mu}F^{\mu\nu a}D_{\rho}F^{\rho}{}_{\nu}{}^{b} + a_{2}\kappa^{\alpha\beta}D_{\mu}\phi_{\alpha}D^{\mu}\phi_{\beta}\,\, ,
\label{intro}
\eea
where  $\mu, \nu,\dots=0,\dots, D-1$ are space-time indices, $a,b,\dots$, are indices in the adjoint representation of the gauge group and $\alpha,\beta\dots$, are indices in the real representation of the same group. The fundamental fields of the theory are a gauge field $A_{\mu}{}^{a}$, and a scalar field $\phi_{\alpha}$. The coefficients $a_1$,$a_2$ are real and depend on the space-time dimension through
\bea
a_1 = -2\left(\frac{D-3}{D-2}\right) \, , \quad
a_2 & = & \frac{a_1}{2(D-1)}\, .
\label{coeffquadratic}
\eea

The authors of \cite{EA} showed that \eqref{intro} is related to a higher-derivative double theory (HDDT), which reproduces the quadratic Weyl gravity Lagrangian once the strong constraint is imposed and the matter contributions from the $b$-field and the dilaton are eliminated. The main goal of this paper is to understand how to classify this HDDT within the construction presented in \cite{MN}, which provides a classification of all possible four-derivative theories in the double geometry. Since the HDDT does not deform Lorentz symmetry or infinitesimal diffeomorphisms, its parametrization should correspond to bosonic supergravity, up to four-derivative terms, written in the Metsaev–Tseytlin form \cite{MT}, modulo field redefinitions. In particular, the Metsaev–Tseytlin action does not contain quadratic contributions from the $b$-field and the dilaton, and therefore such terms in the HDDT must be removed by field redefinitions.

In this work, we explicitly prove that all matter contributions from the $b$-field and the dilaton coming from the off-shell double copy map of \eqref{intro} can indeed be eliminated by field redefinitions (Section 3). We then show that uplifting this formulation to a double field theory with higher-derivative terms requires deforming DFT$+$ by terms proportional to the leading-order equations of motion (Section 4). The findings presented here provide a starting point for several potential research directions (outlined in the discussion section), contributing toward the long-term goal of constructing $\alpha'$-corrections within the DC framework. In the next section, we begin by briefly reviewing \cite{LMR, EA}, which serve as the foundation for the rest of our work.

\section{The Off-Shell Double Copy Beyond Leading Order}

In \cite{LMR}, a method was proposed for introducing higher-derivative corrections such that the resulting double theory exhibits a connection to conformal symmetry. The construction begins with the higher-derivative gauge Lagrangian  
\be
{\cal L} = \frac{1}{2}\kappa_{ab}D_{\mu}F^{\mu\nu a}D_{\rho}F^{\rho}{}_{\nu}{}^{b} \, ,
\label{DFDFold}
\ee
to which the double copy prescription of \cite{HohmDC}, relating the gauge field $A_{\mu}{}^{a}(x)$ with the gravitational object $e_{\mu\bar{\nu}}(x,\bar{x})$, is applied. Under this mapping, the gauge theory \eqref{DFDFold} is uplifted to a higher-order, double-geometric framework,  
\be
\left[\textrm{DFDF}(A_{\mu}{}^{a}) \xrightarrow{DC} \textrm{CDFT}(e_{\mu\bar{\nu}},\Phi)\right]^{(2,3)} \, \nn ,
\ee
whose degrees of freedom match those of perturbative DFT. In the pure gravitational limit\footnote{When the Kalb--Ramond field and the dilaton are non--zero, matter contributions must also be taken into account.}  
(pg: $g_{\mu\nu}=h_{\mu\nu}$, $b_{\mu \nu}=\varphi=0$), the theory reduces to quadratic Weyl gravity,  
\be
\left[\textrm{CDFT}(e_{\mu\bar{\nu}},\Phi) \xrightarrow{pg} \textrm{Weyl gravity}(h_{\mu \nu})\right]^{(2)} \, \nn ,
\ee
subject to the gauge condition
\be
\Box h - \partial_{\mu}\partial_{\nu}h^{\mu\nu}=0 \, .
\label{gaugefixingold}
\ee

The need for this gauge fixing highlights a limitation: not all terms of the quadratic Weyl Lagrangian are reproduced, preventing a fully gauge independent and non-perturbative formulation. To overcome this restriction, the authors of \cite{EA} introduced a charged scalar field $\phi_{\alpha}$. Their gauge theory takes the form  
\bea
{\cal L} & = & a_{1}\kappa_{ab}D_{\mu}F^{\mu\nu a}D_{\rho}F^{\rho}{}_{\nu}{}^{b} + a_{2}\kappa^{\alpha\beta}D_{\mu}\phi_{\alpha}D^{\mu}\phi_{\beta} + a_{3}f_{abc}F_{\mu}{}^{\nu a}F_{\nu}{}^{\lambda b}F_{\lambda}{}^{\mu c}\, \nn \\
& & + a_{4}C^{\alpha}{}_{ab}\phi_{\alpha}F_{\mu\nu}{}^{a} F^{\mu\nu b} + a_{5}d^{\alpha\beta\gamma}\phi_{\alpha}\phi_{\beta}\phi_{\gamma} \, ,
\label{fullgauge_sect_2}
\eea
where the $a_{i}$ are real constants to be fixed. The scalar transforms under the same gauge group as the vector field, but in a real representation labeled by indices $\alpha,\beta,\dots$, and it couples directly to the gauge sector. This construction, inspired by \cite{DFJ1,DFJ2}, is related to the double copy of Weyl gravity at the level of scattering amplitudes and the off-shell realization of the double copy procedure give rise to a geometrical framework which embeds Weyl gravity plus $b$-field and dilaton (matter) contributions, naturally incorporated by the generalized geometry of double field theory. 

Focusing on the quadratic contributions (with $a_i$ fixed by \ref{coeffquadratic}), one obtains  
\bea
{\cal L} & = & a_{1}\kappa_{ab}D_{\mu}F^{\mu\nu a}D_{\rho}F^{\rho}{}_{\nu}{}^{b} + a_{2}\kappa^{\alpha\beta}D_{\mu}\phi_{\alpha}D^{\mu}\phi_{\beta}\,\, ,
\label{fullgauge_sect_2}
\eea
whose double copy at quadratic order yields a perturbative Higher--Derivative Double Theory (HDDT),  
\be
\left[a_1 (DF)^2 + a_2 (D\phi)^2 \xrightarrow{DC} \textrm{pert. HDDT}\right]^{(2)}\, .
\label{map}
\ee
Its field content consists of a generalized frame $e_{\mu\bar{\mu}}(x,\tilde{x})$ and a generalized dilaton $\Phi(x,\tilde{x})$, obtained by identifying $A_{\mu}{}^{a}$ and $\phi_{\alpha}$ as in \cite{LMR, EA}. In particular, \cite{EA} extended this analysis by including interaction terms in the Lagrangian \eqref{fullgauge_sect_2}, demonstrating the existence of HDDT beyond quadratic order.  

\section{Higher-Derivative Double Copy in Supergravity}

The quadratic DC Lagrangian obtained from \eqref{map}, after performing all identifications, is given by \cite{EA}
\bea
\label{DFDF_DFT}
S_{\rm{DC}}^{(2)} & = & - \frac{1}{2}\int d^{D}x d^{D}\bar{x}\Big[a_{1}\Big(\Box e^{\mu\bar{\nu}}\Box e_{\mu\bar{\nu}} - \Box e^{\mu\bar{\nu}}\partial_{\mu}\partial^{\rho}e_{\rho\bar{\nu}} - \Box e^{\mu\bar{\nu}}\bar{\partial}_{\bar{\nu}}\bar{\partial}^{\bar{\sigma}}e_{\mu\bar{\sigma}} + \partial^{\mu}\bar{\partial}^{\bar{\nu}}e_{\mu\bar{\nu}}\partial^{\rho}\bar{\partial}^{\bar{\sigma}}e_{\rho\bar{\sigma}}\Big)\, \nn \\ 
& & \ \ \ \ \ \ \ \ \ \ \ \ \ \ \ \ \ \ \ \ - 2a_{2}\Big(\partial_{\mu}\partial_{\bar{\nu}}e^{\mu\bar{\nu}} + 2\Box\Phi\Big)^{2}\Big]\, .
\eea
While its main feature is to capture Weyl gravity in the pure gravity limit, this Lagrangian also produces $b$-field contributions since $e_{\mu \nu}=h_{\mu \nu} + b_{\mu \nu}$ after identifying the pair of space-time indices, and dilaton contributions which remained unexplored in \cite{EA}. The ambiguous nature of these contributions under field redefinitions is crucial for understanding how to uplift this setup to DFT$+$. For this reason, in the next subsections we analyze separately the quadratic $b$-field and dilaton contributions.

\subsection{$b$-field Quadratic Contributions}

It is easy to show that terms mixing $b_{\mu \nu}$ and $h_{\mu \nu}$ cancel, and the remnant contributions are given by
\bea
L^{(2)}_{b} = - \frac{a_1}{2}(\Box b^{\mu \nu} \Box b_{\mu \nu} - 2 \Box b^{\nu \mu} \partial_{\mu} \partial^{\rho} b_{\nu \rho}) \,  .
\eea
Defining $\bar h_{\mu \nu \rho}= 3 \partial_{[\mu} b_{\nu \rho]}$, the previous Lagrangian can be written as
\bea
L^{(2)}_b = \frac{a_1}{6} \Box \bar h^{\mu \nu \rho} \bar h_{\mu \nu \rho} \, . 
\label{bcontribution}
\eea

Comparing with the four-derivative contributions to bosonic string theory in the Metsaev–Tseytlin formalism \cite{MT},
\bea
L^{(1)} & = & \frac{\alpha'}{4} \Big[  \textrm{Riem}^2 - \frac12 H^{\rho \mu}{}_{\nu} H_{\mu}{}^{\alpha \lambda} R^{\nu}{}_{\rho \alpha \lambda} + \frac{1}{24} H^4 - \frac18 H^2_{\mu \nu} H^{2 \mu \nu} \Big]  \, \, 
\label{MT}
\eea
we see that there are no quadratic contributions in the $b$-field. Therefore, the contribution obtained in \eqref{bcontribution} must correspond to a field redefinition. Indeed, let us consider 
\bea
\label{b-redef}
b_{\mu \nu} = b^{(0)}_{\mu \nu} + c_1 \nabla_{\rho} \bar h^{\rho}{}_{\mu \nu} 
\eea
with $c_1$ an arbitrary coefficient. Let's apply this field redefinition on the leading order term $T_{H}=-\frac{1}{12} H^2$ to quadratic order in perturbations. The induced four-derivative contribution is given by
\bea
T_{\textrm{induced}} = - \frac{c_1}{2} \partial_{\mu}(\nabla_{\lambda} \bar h^{\lambda}{}_{\nu \rho}) \bar h^{\mu \nu \rho} \, .
\eea 
Since we are considering only quadratic contributions, the covariant derivative reduces to an ordinary derivative $\nabla \rightarrow \partial$, and therefore we can write the previous expression as
\bea
T_{\textrm{induced}} = -\frac{c_1}{6} \Box \bar h_{\mu \nu \rho} h^{\mu \nu \rho} \, .
\label{redefb}
\eea
By fixing $c_1= -a_1$, we can interpret the DC theory \eqref{map} as Weyl gravity plus an ambiguous $b$-field contribution, which can be removed by the field redefinition \eqref{b-redef}.

\subsection{Dilatonic Quadratic Contributions}

A similar analysis applies to the dilaton. The quadratic dilaton terms in \eqref{map} are
\bea
L^{(2)}_{\phi}= - 8 a_2 (\Box h - \partial_{\mu} \partial_{\nu} h^{\mu \nu} + \Box \phi) \Box \phi \, .
\eea
Since the dilaton does not contribute to the quadratic structure of the $\alpha'$-corrections, these terms must also be removable by field redefinitions. Consider
\bea
\phi = \phi^{(0)} + c_2 (\Box h - \partial_{\mu} \partial_{\nu} h^{\mu \nu} + \Box \phi) \, .
\eea
Inspecting the two-derivative term $4 \partial_{\mu} \phi \partial^{\mu} \phi$, one finds that the desired contribution is induced by setting $c_2=-2 a_2$. Therefore, both the $b$-field and dilaton contributions in \eqref{map} are ambiguous at quadratic order, leaving Weyl gravity as the only unambiguous sector.  

\section{DFT$+$ Behind Double Copy Maps} 
\label{HDFT}

Now, we will explore the possibility of using the off-shell DC map to generate perturbative, four-derivative corrections to the bosonic string. To this end, we begin by considering DFT+ \cite{Hohm:2014xsa,MN}, a well-known T-duality invariant theory whose four-derivative corrections are related to the closed bosonic string theory upon parametrization. The corresponding action is given by \eqref{biparametric}, up to field redefinitions. As emphasized in \cite{Hohm:2014xsa}, which first introduced the name DFT+, this framework consistently incorporates $\alpha'$-corrections to the low-energy effective action of the bosonic string while preserving the key symmetries of Double Field Theory. When the strong constraint is imposed, the theory reduces to the standard NS-NS supergravity Lagrangian, along with the first-order $\ap$ corrections of the closed bosonic string. This reduced action corresponds to the result obtained by applying the DC prescription to a theory that includes Yang-Mills (YM) together with \eqref{fullgauge_sect_2}, suggesting the relation. 

The starting point is a quadratic gauge Lagrangian $L(\tilde A,A)$ invariant under two different gauge groups, whose off-shell double copy is $L_{\textrm{DFT}+}$. Both gauge fields are mapped to a single gravitational field, $\tilde A(x)\rightarrow e(x,\bar x)$ and $A(x)\rightarrow e(x,\bar x)$, with some freedom in the identification of the Cartan-Killing metrics $\tilde \kappa$, $\kappa$ and structure constants $\tilde f$ and $f$
\be
\left[YM(\tilde F) + a_1(DF)^{2} + a_2(D\phi)^{2} \xrightarrow{DC} {\rm{{pert.}}} \ \textrm{DFT+}\right]\, .
\label{mapextended}
\ee
While one might be tempted to identify the gauge groups such that $\tilde F = F$, the term with $a_1$ would be proportional to the leading-order equation of motion and is therefore ambiguous due to field redefinitions.\footnote{We thank O. Hohm and F. Diaz-Jaramillo for this observation.} After the double-copy procedure, we find that this particular feature results in a DFT$^+$ with extra terms proportional to the leading-order equation of motion, which in turn recovers the quadratic $\alpha'$ corrections to the Weyl gravity Lagrangian (see equation \ref{biparametric}).

\subsection{Inducing Weyl Gravity From DFT$+$}

We construct the non-perturbative DFT+ theory (up to field redefinitions) by considering ${\rm{O}}(D,D)$ multiplets in a double geometry, following the standard approach to incorporating higher-derivative terms into DFT. The dimension of the double space is $2D$, so coordinates are defined as $X^{M}=(x^{\mu},\tilde{x}_{\mu})$, with $M,N,\dots$ indices in the fundamental representation of the duality group. The coordinates $\tilde{x}_{\mu}$ are dual coordinates and are removed by imposing the strong constraint,
\be
\partial_{M} (\partial^{M} \star) = (\partial_{M} \star) (\partial^{M} \star) = 0 \, ,
\ee
where $\star$ denotes arbitrary generalized fields/parameters or products of them. The fundamental fields are a generalized frame $E_{M}{}^{A}$ and a generalized dilaton $d$. These fields transform with respect to generalized diffeomorphisms and double Lorentz transformations as
\be
\delta_{\hat \xi,\Lambda} E_{M}{}^{A} = {\cal L}_{\hat \xi} E_{M}{}^{A} + E_{M}{}^{B} \Lambda_{B}{}^{A} \, ,  \qquad
\delta_{\hat \xi} d = \hat \xi^{N} \partial_N d - \frac12 \partial_{M} \hat \xi^{M} \, , 
\ee
where ${\cal L}_{\hat \xi}$ is the generalized Lie derivative. The generalized frame satisfies,
\be
E_{M A} {\cal H}^{A B} E_{N B} = {\cal H}_{M N} \, , \quad E_{M A} \eta^{A B} E_{N B} = \eta_{M N} \, ,
\ee
where $\eta^{AB}$ and ${\cal H}^{AB}$ are double Lorentz invariant metrics, $\eta_{M N}$ is the ${\rm{O}}(D,D)$ invariant metric, and ${\cal H}_{M N}$ is known as the generalized metric. Using these metrics we can define the (flat) projectors
\be
P_{AB} = \frac{1}{2}\left(\eta_{AB} - {\cal  H}_{AB}\right) \, , \quad \ov{P}_{AB} = \frac{1}{2}\left(\eta_{AB} + {\cal H}_{AB}\right)\ , \nn
\ee
acting on arbitrary double-Lorentz vectors, leads to $P_{A}{}^{B}V_{B} = V_{\un{A}}$ and $\ov{P}_{A}{}^{B}V_{B} = V_{\ov{A}}$.

At this point it is necessary to introduce the generalized fluxes 
\be
F_{A B C} = 3 \partial_{[A} E^{M}{}_{B} E_{|M| C]} \, , \qquad
F_{A} = \sqrt{2}e^{2d}\partial_{M}\left(e^{-2d}E^{M}{}_{A}\right)\, , 
\ee
with the flat derivative defined as $\partial_{A}=\sqrt{2} E^M{}_{A} \partial_{M}$. The action of DFT$+$ is given by 
\be
{\cal S}_{\textrm{DFT+}} = \int d^{2D}X e^{-2d} \Big({\cal R} -{\cal R}^{(-)} 
- {\cal R}^{(+)} + \frac{1}{2(D-2)} {\cal R}_{M N}{\cal R}^{M N} - \frac{1}{2(D-2)(D-1)} {\cal R}^2 \Big)\, , 
\label{biparametric}
\ee
with $\cal R$ the generalized Ricci scalar, ${\cal R}_{M N}$ the generalized Ricci tensor and
\bea
{\cal R}^{(+)} & = & - \frac{1}{2}\Big[(\partial_{\un{A}}\partial_{\un{B}}F^{\un{B}}{}_{\ov{CD}})F^{\un{A}\ov{CD}} + (\partial_{\un{A}}\partial_{\un{B}}F^{\un{A}}{}_{\ov{CD}})F^{\un{B}\ov{CD}} + 2(\partial_{\un{A}}F_{\un{B}}{}^{\ov{CD}})F^{\un{A}}{}_{\ov{CD}}F^{\un{B}} \nn \\ & & + (\partial_{\un{A}}F^{\un{A}\ov{CD}})(\partial_{\un{B}}F^{\un{B}}{}_{\ov{CD}})\Big.\, 
 + (\partial_{\un{A}}F_{\un{B}}{}^{\ov{CD}})(\partial^{\un{A}}F^{\un{B}}{}_{\ov{CD}}) + 2(\partial_{\un{A}}F_{\un{B}})F^{\un{B}}{}_{\ov{CD}}F^{\un{A}\ov{CD}} \nn \\ & & + (\partial_{\ov{A}}F_{\un{B}\ov{CD}})F_{\un{C}}{}^{\ov{CD}}F^{\ov{A}\un{BC}}  - (\partial_{\un{A}}F_{\un{B}\ov{CD}})F_{\un{C}}{}^{\ov{CD}}F^{\un{ABC}}\, + 2(\partial_{\un{A}}F^{\un{A}}{}_{\ov{CD}})F_{\un{B}}{}^{\ov{CD}}F^{\un{B}} \nn \\ & & - 4(\partial_{\un{A}}F_{\un{B}}{}^{\ov{CD}})F^{\un{A}}{}_{\ov{CE}}F^{\un{B}\ov{E}}{}_{\ov{D}} + \frac{4}{3}F^{\ov{E}}{}_{{\un{A}}\ov{C}}F_{{\un{B}}\ov{ED}}F_{\un{C}}{}^{\ov{CD}}F^{\un{ABC}} + F^{\un{B}}{}_{\ov{CD}}F_{\un{A}}{}^{\ov{CD}}F_{\un{B}}F^{\un{A}}\, \nn \\
& &  \Big. + F_{\un{A}}{}^{\ov{CE}}F^{}_{\un{B}\ov{ED}}F^{\un{A}}{}_{\ov{CG}}F^{\un{B}\ov{GD}} - F_{\un{B}}{}^{\ov{CE}}F^{}_{\un{A}\ov{ED}}F^{\un{A}}{}_{\ov{CG}}F^{\un{B}\ov{GD}} - F_{\ov{A}\un{BD}}F^{\un{D}}{}_{\ov{CD}}F_{\un{C}}{}^{\ov{CD}}F^{\ov{A}\un{BC}}\Big]\, , \nn \\ 
\label{Rplus}
\eea
\normalsize
is a higher-derivative combination \cite{MN} that can be constructed by extending to the heterotic duality group and applying the generalized Bergshoeff-de Roo identification \cite{gbdr1}, while ${\cal R}^{(-)}= {\cal R}^{(+)}[P\leftrightarrow \ov P]$. The four-derivative extra terms depending on ${\cal R}$ and ${\cal R}_{\ov{A}\un{B}}$ are contributions that are typically not considered in higher-derivative DFT, as they are proportional to the leading-order equations of motion and can therefore be absorbed through field redefinitions.

DFT+ is invariant under global ${\rm{O}}(D,D)$ transformations, local generalized diffeomorphisms, and double Lorentz transformations. The latter are deformed via a Green-Schwarz-type mechanism,
\be
\delta^{(1)}E_{M}{}^{\ov{A}} = - F_{\un{B}}{}^{\ov{CD}}\partial^{\ov{A}}\Lambda_{\ov{CD}}E_{M}{}^{\un B}\, , \qquad   
\delta^{(1)}E_{M}{}^{\un{A}} = F_{\un{A}}{}^{\ov{CD}}\partial_{\ov{B}}\Lambda_{\ov{CD}}E_{M}{}^{\ov{B}}\, .
\label{GS}
\ee
While one might be tempted to break Lorentz invariance and define $S_{\textrm{HDDT}}$ as consisting only of the four-derivative terms in (\ref{biparametric}), thereby recovering the double copy map (\ref{map}) and restoring ordinary Lorentz invariance upon parametrization, this proposal for the HDDT action does not reduce to Weyl gravity in the pure gravitational limit. The inclusion of the generalized Ricci scalar is crucial due to higher-derivative field redefinitions at the supergravity level.

The parametrization of the DFT+ degrees of freedom follows the same form as in standard DFT
\be
E^{ M}{}_{ A} \ = \
\frac{1}{\sqrt{2}}\left(\begin{matrix}-{ e}_{\mu \underline a}-b_{ \rho\mu} { e}^{\rho }{}_{\underline a} &  { e}^{\mu }{}_{\underline a} \, , \\ 
e_{\mu \overline a}-b_{\rho \mu}{} e^{\rho }{}_{\overline{a}}& e^\mu{}_{\overline a} \end{matrix}\right)  \, , \qquad
e^{-2d} = \sqrt{-\tilde g} e^{-2 \tilde \phi} \, ,
\ee
where $e_{\mu \underline a}$ and $e_{\mu \overline a}$ are two vielbein generating the same $D$-dimensional metric $\tilde g_{\mu\nu}$. As mentioned previously, it is necessary to perform the following field redefinition,
\be
\label{metricredef}
\tilde{g}_{\mu \nu} = g_{\mu \nu} - \frac{\ap}{2}  \Omega^{(-)}_{\mu ab} \Omega^{(-)}_{\nu}{}^{ab} - \frac{\ap}{2} \Omega^{(+)}_{\mu ab} \Omega^{(+)}_{\nu}{}^{ab}\, ,
\ee
to eliminate the anomalous transformation of the metric under Lorentz transformations. Additionally, the dilaton must be redefined to ensure that the integration measure remains invariant, i.e., $\sqrt{-\tilde g}e^{-2\tilde{\phi}}=\sqrt{-g}e^{-2{\phi}}$. In \eqref{metricredef} we include the torsionful spin connection
\be
\Omega^{\pm}_{\mu ab} = w_{\mu ab} \pm \frac{1}{2}H_{\mu ab}\, , 
\ee
where $w_{\mu ab}$ is the spin connection, and $H_{\mu ab}$ is the field strength of the Kalb-Ramond field.

The low energy effective action that describes the previous theory once the duality group is broken is given by
\be
S =   \int d^{D}x  \sqrt{-g} e^{-2\phi} \left[ {\cal L}^{(2)} -\frac{\alpha'}{4} \left( \textrm{Riem}^2 - \frac{4}{D-2} R_{\mu \nu} R^{\mu \nu} + \frac{2}{(D-2)(D-1)} R^2  + {\cal L}^{(4)}_{\textrm{matter}}(b,\phi) \right) \right]\, ,  
\label{alphasugra}
\ee
with ${\cal L}^{(2)}$ the NS-NS sector of supergravity. The four-derivative terms of this Lagrangian reproduces Weyl gravity in the pure gravitational limit (p.g.: $b_{\mu \nu}=\phi=0$), and by setting $\alpha'=-4$ the action reduces to,
\bea
S|_{\textrm{p.g.}}= \int d^{D}x  \sqrt{-g} \, \left(R + C_{\mu \nu \rho \sigma} C^{\mu \nu \rho \sigma} \right)\, , 
\label{Weyl}
\eea
where $C_{\mu \nu \rho \sigma}$ is the Weyl tensor. This is the theory obtained from applying the DC prescription on the gauge side of \eqref{mapextended} and hence, the confirmation of the existence of a DC map between that gauge theory and DFT+.

\subsection{Elimination of ${\cal L}^{(4)}_{\textrm{matter}}(b,\phi)$}
As a final step, we show that ${\cal L}^{(4)}_{\textrm{matter}}(b,\phi)$ can be eliminated by field redefinitions, in agreement with the section 2.

\subsubsection{Analysis of ${\cal R}_{M N} {\cal R}^{M N}$ Term}
The $b$-field and dilaton contributions coming from this term are given by
\bea
{\cal R}_{M N} {\cal R}^{M N}|_{b,\phi} & = & 2 \Big[2 R_{\mu \nu}(- \frac14 H^{\mu \rho \sigma} H^{\nu}{}_{\rho \sigma} + 2 \nabla^{\mu} \nabla^{\nu} \phi) \nn \\ && + (- \frac14 H^{\mu \rho \sigma} H^{\nu}{}_{\rho \sigma} + 2 \nabla^{\mu} \nabla^{\nu} \phi) (- \frac14 H_{\mu}{}^{\epsilon \tau} H_{\nu \epsilon \tau} + 2 \nabla_{\mu} \nabla_{\nu} \phi)\Big] \nn \\ && + 2 (\frac12 \nabla^{\rho} H_{\rho \mu}{}^{\nu} - H_{\rho \mu}{}^{\nu} \nabla^{\rho}\phi)(\frac12 \nabla^{\sigma} H_{\sigma}{}^{\mu}{}_{\nu} - H_{\sigma}{}^{\mu}{}_{\nu}  \nabla^{\sigma}\phi) \, . 
\eea
The quadratic contributions take the form
\bea
{\cal R}_{M N} {\cal R}^{M N}|^{(2)}_{b,\phi} & = & 8 R_{\mu \nu}\partial^{\mu} \partial^{\nu} \phi + 8 \partial^{\mu} \partial^{\nu} \phi \partial_{\mu} \partial_{\nu} \phi + \frac12 \partial^{\rho} \bar h_{\rho \mu}{}^{\nu} \nabla^{\sigma} \bar h_{\sigma}{}^{\mu}{}_{\nu} \, . 
\label{extra1}
\eea
We now show that all these terms can be removed by appropriate field redefinitions of the metric, the $b$-field and the dilaton. The term proportional to $R_{\mu\nu}$ can be eliminated by a covariant redefinition of the metric
\bea
g_{\mu \nu} = g^{(0)}_{\mu \nu} + 8 \nabla_{\mu} \nabla_{\nu} \phi \, .
\eea
At quadratic order this induces
\bea
L^{(2)}_{\textrm{induced}} = 8 (\Box \phi R - R_{\mu \nu} \partial^{\mu} \partial^{\nu} \phi) \, .
\eea
The first term can be absorbed into a dilaton redefinition, while the second exactly cancels the first term in \eqref{extra1}. The second term in \eqref{extra1} can be rewritten, using partial integration, as $8 \Box \phi \Box \phi$ and this structure can be removed by a dilaton redefinition. Finally, the third term can be brought to the form of \eqref{redefb}, and thus eliminated by the same $b$-field redefinition used earlier.

\subsubsection{Analysis of ${\cal R}^2$ Term}

The $b$-field and dilaton contributions from this term are
\be
{\cal R}^{2}|_{b \phi} = 2 R (-4 \nabla_{\mu} \phi \nabla^{\mu}\phi + 4 \nabla_{\mu} \nabla^{\mu}\phi - \frac{1}{12} H^2) + (-4 \nabla_{\mu} \phi \nabla^{\mu}\phi + 4 \nabla_{\mu} \nabla^{\mu}\phi - \frac{1}{12} H^2)^2 \, ,
\ee
which at quadratic order yields ${\cal R}^{2}|^{(2)}_{b \phi} = (8 R + 16 \Box \phi) \Box \phi$. These terms can again be eliminated by a dilaton redefinition of the form
\be
\phi = \phi^{(0)} + \tfrac14 (8 R + 16 \Box \phi) \, ,
\ee
closely following the strategy employed in Section~2.

\section{Discussion}

In this work, we have explored the role of the Double Copy approach as a tool for constructing higher-derivative corrections in the context of Double Field Theory. Our analysis demonstrates that applying the off shell DC prescription \cite{HohmDC, LMR, EA} to the Yang-Mills Lagrangian, together with higher-derivative terms from gauge theory, leads to a perturbative structure consistent with previous studies on higher-derivative corrections, of the form DFT+HDDT. Furthermore, we show that the DFT+HDDT structure is connected to an exact formalism given by DFT+ theory, including four-derivative corrections invariant under global ${\rm{O}}(D,D)$ transformations and local generalized diffeomorphisms. These corrections align with the $\alpha'$ expansion of the closed bosonic string when properly parameterized. Notably, the action contains contributions essential for maintaining Lorentz invariance and recovering Weyl gravity in the pure gravitational limit, remarking the importance of higher-derivative field redefinitions at the supergravity level.

Our results demonstrate that the higher-derivative structure of DFT$+$ can be recovered from an off-shell double copy construction starting from a suitable gauge theory containing both Yang–Mills and four-derivative $(DF)^2$ and $(D\phi)^2$ operators. At quadratic order, the DC map reproduces a perturbative Higher-Derivative Double Theory (HDDT) whose pure gravitational sector is given by Weyl gravity, while the $b$-field and dilaton contributions can be removed via field redefinitions. This is consistent with the known structure of the first $\alpha'$-corrections in the bosonic string, where quadratic $b$-field and dilaton terms are absent in the Metsaev–Tseytlin formalism and can be shifted away. At cubic order, compatibility between HDDT and DFT$+$ is plausible but not automatic: producing the $H^{2}\text{Riem}$ term of the Metsaev–Tseytlin action may require additional gauge structure. While Weyl gravity follows directly from the $(DF)^2$ operator (plus scalar fields), the consistent coupling of the $b$-field and dilaton could demand further extensions. 

On the other hand, constructing a Conformal Field Theory (CDFT) within the double geometry, as outlined in \cite{Rasim}, is limited by double Lorentz transformations. To the best of our knowledge, the only higher-derivative combination in the bi-parametric extension given in \cite{MN} that can be expressed in terms of the generalized metric is provided by the DFT- or HSZ theory \cite{HSZ,HSZ2}, whose parameterization does not lead to $\text{Riem}^2$ contributions. We do not rule out the possibility that CDFT could include additional non-covariant contributions, potentially incorporating non-covariant terms to resolve the lack of invariance within the generalized frame formalism (both in the bosonic or heterotic cases). We leave this question open for future investigation.

A key conceptual outcome of this work is the explicit realization that DFT$+$ arises naturally as the gravitational side of a DC map involving higher-derivative gauge theories. While DFT$+$ was originally motivated by the systematic inclusion of $\alpha'$-corrections into a doubled, T-duality covariant setting, we find that it also serves as the completion of the double copy of a particular higher-derivative gauge system related to Weyl gravity. This perspective strengthens the connection between color–kinematics duality and duality-invariant gravitational theories, and suggests that the DC framework may provide a systematic route to higher-curvature corrections that go beyond standard supergravity methods.

Several directions for future research follow from our analysis: 
\begin{itemize}

\item \textbf{New Double Copy Realizations:} The connection established here between the DC map and DFT+ provides a solid foundation for further exploring links between gauge theories and T-duality-invariant frameworks. A natural step is to relax the covariant constraints of our model and compare results at the level of amplitudes. We expect agreement with \cite{Newref1,Newref2} upon imposing the equations of motion. This may require extending the minimal set of identifications presented in this work. Another possible direction is to match our construction with the HSZ theory \cite{HSZ,HSZ2}, in line with the bi-parametric model discussed in \cite{MN}. Interpolating between DFT$+$ (bosonic DFT with higher-derivative terms) and DFT$-$ may give access access to perturbative heterotic $\alpha'$-corrections. Similarly, comparisons with the HSZ model would be relevant in the context of six-derivative terms ($\text{Riem}^3$) in bosonic supergravity, which partially overlap with the corrections found in the bosonic string and could be further explored in light of \cite{Newref3}.

\item \textbf{Conformal Double Field Theory:} The findings of this paper are highly relevant for refining the identifications within the DC procedure described in \cite{EA}. While the original objective was to identify the gauge field necessary for constructing a CDFT, we now propose an alternative approach: incorporating the Yang-Mills Lagrangian and identifying the fields that match the $\alpha'$ structure of closed bosonic string theory. Although a complete construction of CDFT using the techniques presented in this paper is not yet possible, recent advancements have been made in this direction \cite{Rasim}. It is important to clarify that CDFT is not defined solely through the higher-derivative structure of DFT+, but the methods developed in this work could serve as a solid framework for building the CDFT Lagrangian by considering a non-covariant extension of the bi-parametric family of theories outlined in \cite{MN}.

\item \textbf{Single and Zeroth Copies in Higher-Derivative Theories:} The study of the single and zeroth copies of perturbative DFT was initially addressed in \cite{KL} and further developed in \cite{KSextra1, KSextra2, KSextra3, KSextra4, KSextra5, KSextra6}. The perturbative ansatz used to link DFT and Yang-Mills dynamics via the single and zeroth copies is a generalized version of the Kerr-Schild ansatz. The results of this work now open the possibility of investigating higher-derivative corrections to the generalized Kerr-Schild ansatz, connecting DFT+ to the higher-derivative gauge theory described in \eqref{fullgauge_sect_2}, thereby extending the framework in \cite{KSextra3}.
\end{itemize}

\subsection*{Acknowledgements}

E.L. is supported by the SONATA BIS grant 2021/42/E/ST2/00304 from the National Science Centre (NCN), Poland. J.A. R. gratefully acknowledges the support provided by Universidad de Buenos Aires (UBA).

\end{document}